\documentclass[useAMS]{mn2e}

\usepackage{graphicx,color,soul}
\usepackage{latexsym}
\usepackage{multirow}
\usepackage{amsmath,amssymb}
\usepackage{epstopdf}
\usepackage[draft=false]{hyperref}

\title[Alfv\'en waves in the process of heating the quiet solar corona]{Estimating the contribution of Alfv\'en waves to the process of heating the quiet solar corona}

\author[J. J. Gonz\'alez-Avil\'es and F. S. Guzm\'an]{J. J. Gonz\'alez-Avil\'es  \thanks{E-mail:javiles@ifm.umich.mx (JJGA)} and F. S. Guzm\'an \thanks{E-mail:guzman@ifm.umich.mx (FSG)}   \\ 	
        Instituto de F\'{\i}sica y Matem\'{a}ticas, Universidad
        Michoacana de San Nicol\'as de Hidalgo. \\ Edificio C3, Cd.
        Universitaria, 58040 Morelia, Michoac\'{a}n,
        M\'{e}xico.
         \\ 
        }

\begin{document}


\date{\today}

\pagerange{\pageref{firstpage}--\pageref{lastpage}} \pubyear{2015}

\maketitle

\label{firstpage}


\begin{abstract}
We solve numerically the ideal MHD equations with an external gravitational field in 2D in order to study the effects of impulsively generated linear and non-linear Alfv\'en waves into isolated solar arcades and coronal funnels. We analyze the region containing the interface between the photosphere and the corona. The main interest is to study the possibility that Alfv\'en waves triggers the energy flux transfer toward the quiet solar corona and heat it, including the case that two consecutive waves can occur. We find that in the case of arcades, short or large, the transferred fluxes by  Alfv\'en waves are sufficient to heat the quiet corona only during a small lapse of time and in a certain region. In the case of funnels the threshold is achieved only when the wave is faster than 10 km/s, which is extremely high. We conclude from our analysis, that Alfv\'en waves, even in the optimistic scenario of having two consecutive Alfv\'en wave pulses, cannot transport enough energy as to heat the quiet corona.
\end{abstract}



\section{Introduction}

Recent observations indicate the existence of Alfv\'en waves \cite{Banerjee_et_al_2007}, kink waves \cite{Aschwanden_et_al_1999,Nakariakov_et_al_199}
and torsional Alfv\'en waves \cite{Jess_et_al_2009} in the solar atmosphere. These observations are based on data provided by current space missions, such as Transition  Region and Coronal Explorer (TRACE), Solar Dynamics Observatory (SDO), Solar Optical Telescope (SOT), X-ray Telescope (XRT) and Swedish Solar Telescope (SST) 
\cite{Banerjee_et_al_2007,Tomczyk_et_al_2007,De_Pontieu_et_al_2007,Okamoto_et_al_2007,Jess_et_al_2009}. Moreover, it has been observed that the solar atmosphere is permeated by magnetic fields, which are organized in several structures, like small-scale magnetic flux tubes, loops, gravitationally stratified and magnetically confined structures (solar magnetic arcades), weakly curved coronal magnetic flux tubes (coronal funnels), and others, and these structures may support the propagation of differents kinds of MHD waves.

We are interested in studying the role that Alfv\'en waves play in the coronal heating problem, but first we mention the physical mechanisms that have been already proposed to study the problem. Theoretical models of coronal heating mechanisms include Direct Current (DC) and Alternating Current (AC) models, which characterize the electromechanic coronal response to the photospheric driver that provides the energy source for heating. The DC models involve  Ohmic dissipation, magnetic reconnection, current cascading and viscous turbulence \cite{Sturrock&Uchida_1981,Parker_1983,Milano_et_al_1999,Gudiksen&Nordlund_2002}, and the AC models involve wave heating by Alfv\'enic resonance \cite{Hollweg_1991}, resonant absorption \cite{Davila_1987,Goossens_et_al_1992,Ofman&Davila_1994,Erdelyi&Goossens_1994}, phase mixing \cite{Parker_1991,De_Moortel_et_al_1999}, current layers \cite{Galsgaard&Nordlund_1996}, MHD turbulence \cite{Matthaeus_et_al_1999,Dmitruk_et_al_2001} and cyclotron resonance \cite{Hollweg_1986,Tu&Marsch_1997}. In addition is it considered that some heating could also be produced by compressional waves (acoustic waves) or shocks \cite{Kuperus_et_al_1981}. Another possible mechanism to heating the solar corona is by magneto-acoustic shock waves generated by chromospheric reconnection \cite{Sturrock_1999,Litvinenko_1999,Sakai_et_al_2000,Ryutova_et_al_2001}.

The theoretical approach is  centered in the analysis of numerical simulations modeling the propagation of Alfv\'en waves in the solar atmosphere, e.g. in \cite{Kudoh&Shibata_1999} numerical simulations have been perfomed to study torsional Alfv\'en waves propagation along an open magnetic flux tube in the solar atmosphere. The results suggest that the hot quiet corona, non-thermal broadening of lines, and spicules are all caused by the non-linear Alfv\'en waves. In \cite{Del_Zanna_et_al_2005}, it has been studied numerically the propagation and evolution of Alfv\'enic pulses in the solar coronal arcades. The results show a propagation of the initially localized pulses within the solar arcade, due to the variations in the  Alfv\'en speed with height, and therefore an efficient damping of the amplitude of the oscillations. In \cite{Murawski&Musielak_2010} small-amplitude Alfv\'en waves are modeled by solving the 1D wave equation and it was found that as a result of cut-off frequencies, these waves are not able to propagate freely into the upper regions of the solar atmosphere, undergoing reflection towards lower layers.

On the other hand, in \cite{Ebadi_et_al_2012} the dissipation of Alfv\'en waves due to phase mixing in a stratified environment of solar spicules has been presented . The results show that in the presence of stratification due to gravity, damping takes place. In \cite{Chmielewski_et_al_2013}, it has been investigated the role of impulsively generated non-linear Alfv\'en waves in the observed non-thermal broadening of some spectral lines in solar coronal holes. The results were that the non-linear Alfv\'en waves with amplitude $A_{v}=50$ km/s are the most likely candidates for the non-thermal broadening of line profiles in the polar coronal hole. In \cite{Chmielewski_et_al_2014a} a 2.5 D numerical model is used to study the propagation and reflection of impulsively generated Alfv\'en waves within a solar magnetic arcade; the results show that the propagation of Alfv\'en waves is affected by the spatial dependence of the Alfv\'en speed, which leads to phase mixing that is stronger for more curved and larger magnetic arcades. In \cite{Chmielewski_et_al_2014b} we found a study of the propagation of impulsively generated linear and non-linear Alfv\'en waves in weakly curved coronal magnetic flux-tubes, or coronal funnels, and it was found that non-linear Alfv\'en waves may carry enough energy as to heat the coronal funnels and power the solar wind that originates in these funnels.

Within all this rich and well investigated scenario we raise again the question of whether Alfv\'en waves are capable of transferring enough energy as to heat the quiet corona. Most of the aforementioned studies focused on this question are based on numerical simulations of a single pulse. In this paper we simulate the propagation of two pulses in order to study the possible effects of the interference in the process of energy transfer to the corona.
 
We study the possible effects of interaction between two impulsively generated Alfv\'en waves in two configurations of magnetic field: i) solar magnetic arcades, that we assume to be isolated and fixed in time so that they may apply to hold on a quiet corona, and ii) coronal funnels. We are interested in the implications of Alfv\'en waves in the heat transfer into the corona. We simulate the interaction by launching two pulses separated by a given interval of time between 0 and 50 s, considering that otherwise the two pulses are considered to be independent.

The paper is organized as follows, in section \ref{sec:modelandNM} we describe the model of solar atmosphere, the numerical methods we use, and the way we measure the energy flux carried by Alfv\'en waves. In section \ref{sec:results}, we present the results of the numerical simulations for various experiments. Finally, in section \ref{sec:conclusions} we present the conclusions. 

\section{Model and numerical methods}
\label{sec:modelandNM}

In this section we describe the equations modeling the plasma, the photosphere-corona interface model, the dimensionless equations solved by our code, the initial conditions of the Alfv\'en wave pulses and the numerical methods used to solve the evolution equations of the system.

\subsection{The system of equations}

We consider a gravitationally stratified solar atmosphere, which is described by a plasma obeying the ideal MHD submitted to an external constant gravitational field. We choose to write down these equations in a conservative form, which is optimal for our numerical methods:

\begin{eqnarray}
&&\frac{\partial\rho}{\partial t} + \nabla\cdot(\rho{\bf v}) = 0, \label{density}\\
&&\frac{\partial(\rho{\bf v})}{\partial t} + \nabla\cdot(\rho{\bf v}\otimes{\bf v}-\frac{{\bf B}\otimes{\bf B}}{\mu_{0}} + {\bf I}p_{t}) = \rho {\bf g},  \label{momentum} \\
&&\frac{\partial E}{\partial t} +\nabla\cdot((E+p_{t}){\bf v}-\frac{{\bf B}}{\mu_{0}}({\bf v}\cdot{\bf B})) ={\rho}{\bf v}\cdot{\bf g}, \label{energy} \\
&&\frac{\partial{\bf B}}{\partial t} +\nabla\cdot({\bf v}\otimes{\bf B} -{\bf B}\otimes{\bf v}) = {\bf 0}, \label{evolB} \\
&&\nabla\cdot{\bf B} = 0, \label{divergenceB} 
\end{eqnarray}

\noindent where $\rho$ is the plasma mass density, $p_{t}=p+\frac{{\bf B}^{2}}{2\mu_{0}}$ is the total (thermal + magnetic) pressure, ${\bf v}$ represents the plasma velocity field, ${\bf B}$ is the magnetic field, $\mu_{0}$ is the vacuum permeability, ${\bf I}$ is the unit matrix and $E$ is the total energy density, that is, the sum of the internal, kinetic, and magnetic energy density

\begin{equation}
E = \frac{p}{\gamma-1} + \frac{\rho{\bf v}^{2}}{2} + \frac{{\bf B}^{2}}{2\mu_{0}}. \label{total_energy}
\end{equation}

\noindent For the fluid we consider an adiabatic index $\gamma=5/3$. The system of equations (\ref{density})-(\ref{divergenceB}) is closed with the ideal gas law

\begin{equation}
p = \frac{k_{B}}{m}\rho T, \label{eos}  
\end{equation}

\noindent where $T$ is the temperature of the plasma, $m=\mu m_{H}$ is the particle mass, which is specified by a mean molecular weight value of $\mu=$1.24 for a fully ionized gas \cite{Chmielewski_et_al_2014a} and $m_{H}$ is hydrogen's mass, $k_{B}$ is Boltzmann's constant. The gravitational source terms on the right hand side of equations (\ref{momentum}) and (\ref{energy}) is given by ${\bf g}=[0,-g,0]$ with magnitude $g=274$ m/$s^{2}$, which holds near the photosphere.

\subsection{Dimensionless MHD equations}

In order to carry out our numerical calculations we rescale the system (\ref{density}-\ref{divergenceB})
according to the following conventions. We choose the length scale $l_{0}$, plasma density scale $\rho_{0}$, and magnetic field scale $B_{0}$ to fix the unit of time in terms of the Alfv\'en speed

\begin{equation}
v_{0}\equiv v_{A,0}\equiv\frac{B_{0}}{\sqrt{\mu_{0}\rho_{0}}} ~~ \Rightarrow ~~ t_{0}\equiv\frac{l_{0}}{v_{0}}. \label{time_scale}
\end{equation}

\noindent Using these basic scale factors $l_{0}$, $B_{0}$, $t_{0}$ we define dimensionless independent variables and associated differential operators as follows

\begin{equation}
\bar{l}\equiv\frac{l}{l_{0}} ~~\Rightarrow ~~\bar{\nabla}\equiv l_{0}\nabla,\quad\bar{t}\equiv\frac{t}{t_{0}} ~~\Rightarrow ~~\frac{\partial}{\partial\bar{t}}\equiv t_{0}\frac{\partial}{\partial t}, \label{ind_variables}
\end{equation}

\noindent and consequently the corresponding dimensionless state variables are defined by 

\begin{eqnarray}
&&\bar{\rho}\equiv\frac{\rho}{\rho_{0}},\quad\bar{{\bf v}}\equiv\frac{{\bf v}}{v_{0}},\quad \bar{p}\equiv\frac{p}{\rho_{0}v_{0}^{2}}, \nonumber \\ 
&&\bar{E}\equiv\frac{E}{\rho_{0}v_{0}^{2}},\quad\bar{{\bf B}}\equiv\frac{{\bf B}}{B_{0}},\quad\bar{{\bf g}}\equiv{\bf g}\left(\frac{l_{0}}{v_{0}^{2}}\right). \label{dep_variables} 
\end{eqnarray}

\noindent Finally the dimensionless equations ruling our system are

\begin{eqnarray}
&&\frac{\partial\bar{\rho}}{\partial\bar{t}} + \bar{\nabla}\cdot(\bar{\rho}{\bf\bar{v}}) = 0, \label{dimensionless_density}\\
&&\frac{\partial{(\bar{\rho}\bf\bar{v}})}{\partial\bar{t}} + \bar{\nabla}\cdot(\rho\bar{{\bf v}}\otimes\bar{{\bf v}}-{\bf\bar{B}}\otimes{\bf\bar{B}} + {\bf I}\bar{p_{t}}) = \bar{\rho}\bar{\bf g},  \label{dimensionless_momentum} \\
&&\frac{\partial\bar{E}}{\partial\bar{t}} + \bar{\nabla}\cdot((\bar{E}+\bar{p}_{t})\bar{\bf v}-\bar{\bf B}(\bar{\bf v}\cdot\bar{\bf B})) = \bar{\rho}\bar{\bf v}\cdot\bar{\bf g}, \label{dimensionless_energy} \\
&&\frac{\partial\bar{\bf B}}{\partial\bar{t}} + \bar{\nabla}\cdot(\bar{\bf v}\otimes\bar{\bf B} - \bar{\bf B}\otimes\bar{\bf v}) = {\bf 0}, \label{dimensionless_evolB} \\
&&\bar{\nabla}\cdot{\bf\bar{B}} = 0. \label{dimensionless_divergenceB} 
\end{eqnarray}

\noindent The dimensionless version of the equation of state (\ref{eos}) is obtained using the definitions of $p$ and $\rho$

\begin{equation}
p =\frac{k_{B}}{m}\rho T ~~ \Rightarrow ~~ \bar{p} = \frac{k_{B}}{v_{0}^{2}m}\bar{\rho}T, \label{dimensionless_eos}
\end{equation}

\noindent and the dimensionless temperature $\bar{T}$ and equation of state are

\begin{equation}
\bar{T}\equiv T\left(\frac{k_{B}}{m v_{0}^{2}}\right),
~~
\bar{p} =\bar{\rho}\bar{T}. \label{dimensionless_eos_1}
\end{equation}

\noindent In this way, the ideal MHD system (\ref{dimensionless_density}-\ref{dimensionless_divergenceB}) is attached to the scale length, magnetic field, density and time factors $l_{0}$, $B_{0}$, $\rho_{0}$ and $t_{0}$. In our simulations we choose the scale factors in cgs units as specified in Table \ref{table:1}. 

\begin{table}
\caption{Units used in this paper}
\centering
\begin{tabular}{c c c c}
\hline\hline
Quantity & cgs units \\ [0.5ex]
\hline
$l_{0}$ & $10^{8}$ cm \\
$v_{0}$ & $10^{8}$ cm$\cdot s^{-1}$ \\
$t_{0}$ & 1 s \\
$\rho_{0}$ & $10^{-15}$ gr$\cdot cm^{-3}$ \\
$B_{0}$ & 14.472 G \\ [1ex]
\hline
\end{tabular}
\label{table:1}
\end{table}

\subsection{Model of the static solar atmosphere}

We specifically center the domain of analysis as one covering part of both the chromosphere and the corona. We consider the atmosphere to be static ($\partial/\partial t=0$) and study the evolution on a finite $xy$ domain, where the $x$ is a horizontal coordinate and $y$ labels height. All the state variables depend on $x$ and $y$, but a non-zero component of the velocity field $v_{z}$ and a non-trivial magnetic field $z$-component $B_{z}$ that vary with $x$ and $y$ are allowed. This kind of model is called a 2.5 dimensional (2.5D) model \cite{Murawski&Musielak_2010,Murawski&Zaqarashvili_2010,Murawski_et_al_2011,Woloszkiewicz_et_al_2014,Chmielewski_et_al_2014a,Chmielewski_et_al_2014b}.

Given the solar atmosphere is in static equilibrium, it should have a force-free and current-free magnetic field given by the condition

\begin{equation}
\nabla\times{\bf B_{e}}={\bf 0} ~~ \Rightarrow ~~
(\nabla\times{\bf B_{e}})\times{\bf B_{e}}={\bf 0},
\label{force_free_B} 
\end{equation}

\noindent where the subscript $_{e}$ stands for equilibrium quantities. The choice of the magnetic field satisfying (\ref{force_free_B}) is general and depends on the physical problem under study. As a result of applying the static equilibrium and the magnetic current-free conditions to equation (\ref{momentum}),  the pressure gradient is balanced by gravity:

\begin{equation}
-\nabla p_{e} +\rho_{e}{\bf g} = {\bf 0}. \label{equilibrium_pressure}
\end{equation}

\noindent In addition, following 
\cite{Murawski&Musielak_2010,Murawski&Zaqarashvili_2010,Murawski_et_al_2011,Woloszkiewicz_et_al_2014,Chmielewski_et_al_2014a,Chmielewski_et_al_2014b} we use the ideal gas law (\ref{eos}) and the fact that hydrostatic equilibrium will only hold along the $y$ direction, so that the equilibrium gas pressure takes the form:

\begin{equation}
p_{e}(y) = p_{ref}\exp\left(-\int_{y_{ref}}^{y}\frac{dy^{\prime}}{\Lambda(y^{\prime})}\right), \label{pe_profile}
\end{equation}

\noindent and the mass density

\begin{equation}
\rho_{e}(y) = \frac{p_{e}(y)}{g\Lambda(y)}, \label{rhoe_profile} 
\end{equation}

\noindent where

\begin{equation}
\Lambda(y) = \frac{k_{B}T_{e}(y)}{mg} \label{scale_pressure}, 
\end{equation}

\noindent is the pressure scale-height, and $p_{ref}$ represents the gas pressure at a given reference level, that we choose to be located at $y_{ref}=10$ Mm, because in the temperature model above, this is the location at which the corona region starts.

\subsubsection{Model for isolated solar magnetic arcades}
\label{sec:shortarcades}

We model an isolated solar magnetic arcade with a magnetic flux-tube model originally developed by \cite{Low1985} in three dimensions. This model considers a magnetic flux function 

\begin{equation}
A(x,y) = \frac{x(y_{ref}-b)^{2}}{(y-b)^{2}-x^{2}}B_{ref}, \label{magnetic_flux}  
\end{equation}

\noindent where $b$ is a constant that determines the vertical location of a magnetic pole, which is set to $b=-5$ Mm and $B_{ref}$ is the magnetic field strength at $y_{ref}$. The magnetic field is thus

\begin{equation}
{\bf B_{e}}(x,y) = \nabla\times(A(x,y){\bf\hat{z}})=\frac{\partial A(x,y)}{\partial y}{\bf\hat{x}} - \frac{\partial A(x,y)}{\partial x}{\bf\hat{y}}, \label{Be} 
\end{equation}

\noindent where specifically the equilibrium magnetic field components $B_{ex}(x,y)$ and $B_{ey}(x,y)$ are

\begin{eqnarray}
B_{ex}(x,y) &=& -\frac{2x(y-b)(b-y_{ref})^{2}}{(x^{2}-(b-y)^{2})^{2}}B_{ref}, \label{Bex}\\
B_{ey}(x,y) &=& -\frac{(x^{2}+(b-y)^{2})(b-y_{ref})^{2}}{(y^{2}-2by-x^{2}+b^{2})^{2}}B_{ref}. \label{Bey}   
\end{eqnarray}

\begin{figure}
\includegraphics[width=7cm]{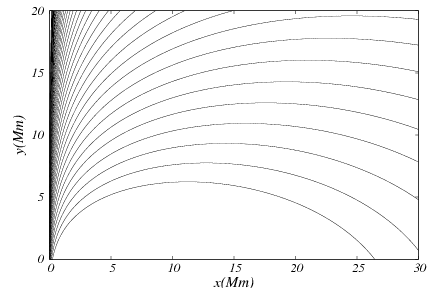}
\includegraphics[width=8cm]{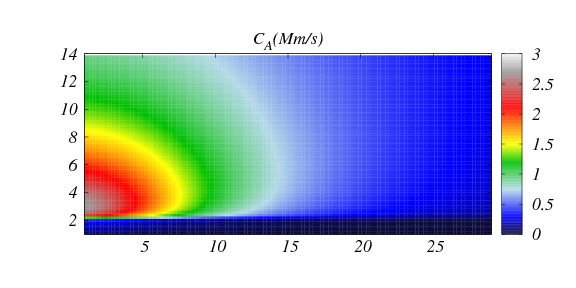}
\caption{\label{fig:Magnetic_lines_CA_xy} (Top) Magnetic field lines for an open arcade with parameters $B_{ref}=14.472$ G, $b=-5$ Mm and $y_{ref}=10$ Mm. (Bottom) The associated Alfv\'en speed in Mm/s.}
\end{figure}

\noindent The value of $B_{ref}$ is choosen such that the Alfv\'en and sound speeds

\begin{equation}
c_{A}(x,y) = \sqrt{\frac{B_{ex}^{2}+B_{ey}^{2}}{\mu_{0}\rho_{e}(y)}}, ~~~
c_{s}(y) = \sqrt{\frac{\gamma p_{e}(y)}{\rho_{e}(y)}}, \label{cA_cs_arcade}
\end{equation}

\noindent satisfy the constraint $c_{A}(0,y_{ref})=10c_{s}(y_{ref})$, because in the solar corona the magnetic pressure is bigger than the fluid pressure. This constraint gives a value of $B_{ref}=14.472$ G. The magnetic field components resulting from equations (\ref{Bex}) and (\ref{Bey}) are displayed in Fig. \ref{fig:Magnetic_lines_CA_xy} (top). Such magnetic configuration corresponds to an isolated asymmetric magnetic arcade as proposed in \cite{Chmielewski_et_al_2014a}. As described in equation (\ref{cA_cs_arcade}) Alfv\'en speed depends on $x$ and $y$, which means that it is non-isotropic as is also shown in Fig. \ref{fig:Magnetic_lines_CA_xy} (bottom). This effect will influence the dynamics of the wave.

\subsubsection{Model for coronal funnels}
\label{sec:funnels}

Following \cite{Chmielewski_et_al_2014b}, we assume the magnetic flux function has only a $\hat{\bf z}$ component and the following expression

\begin{equation}
{\bf A_{e}}(x,y) = \Lambda_{B}B_{ref}\sin\left(\frac{x}{\Lambda_{B}}\right)\exp\left(-\frac{y-y_{ref}}{\Lambda_{B}}\right){\bf\hat{z}}, \label{Magnetix_flux_funnel} 
\end{equation}

\noindent where $B_{ref}$ is again the magnetic field at the reference level $y_{ref}=10$ Mm. Unlike \cite{Chmielewski_et_al_2014b} we use a sin instead of a cosine in order to be consistent with our convention of coordinates. Here $\Lambda_{B}=2L/\pi$ denotes
the magnetic scale-height and $L$ is a half of the magnetic arcade width. In this case we want to model weakly expanding coronal funnels, then we fixed $L=75$ Mm \cite{Chmielewski_et_al_2014b}. For this choice, the magnetic field lines are weakly curved and represent the open and expanding field lines similar to coronal holes.  Finally, the magnetic field reads 

\begin{equation}
{\bf B_{e}}(x,y) = \nabla\times(A(x,y){\bf\hat{z}})=\frac{\partial A(x,y)}{\partial y}{\bf\hat{x}} - \frac{\partial A(x,y)}{\partial x}{\bf\hat{y}}, \label{Be_funnel} 
\end{equation}

\noindent and the equilibrium components $B_{ex}(x,y)$ and $B_{ey}(x,y)$ are explicitly

\begin{eqnarray}
B_{ex}(x,y) &=& -B_{ref}\sin\left(\frac{x}{\Lambda_{B}}\right)\exp\left(-\frac{y-y_{ref}}{\Lambda_{B}}\right) \label{Bex_funnel},\\
B_{ey}(x,y) &=& -B_{ref}\cos\left(\frac{x}{\Lambda_{B}}\right)\exp\left(-\frac{y-y_{ref}}{\Lambda_{B}}\right) \label{Bey_funnel}.
\end{eqnarray}

\noindent On the other hand, in this model the Alfv\'en speed varies only with $y$ because the density does:

\begin{equation}
c_{A}(y) = \sqrt{\frac{B_{ex}^{2}+B_{ey}^{2}}{\mu_{0}\rho_{e}(y)}}=\frac{B_{ref}e^{-\frac{y-y_{ref}}{\Lambda_{B}}}}{\sqrt{\mu_{0}\rho_{e}(y)}}, \label{CA_funnel} 
\end{equation}

\noindent and the equilibrium sound speed is
 
\begin{equation}
c_{s}(y) = \sqrt{\frac{\gamma p_{e}(y)}{\rho_{e}(y)}}. \label{cs_funnel} 
\end{equation}

\begin{figure}
\includegraphics[width=6cm]{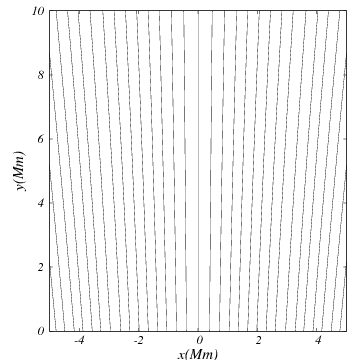}
\includegraphics[width=6cm]{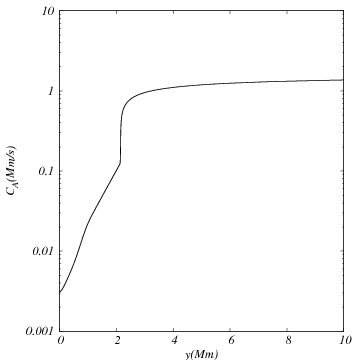}
\caption{\label{fig:Vector_field_funnel_CA} (Top) Magnetic field lines for coronal funnels with parameters $B_{ref}=14.432$ G, $y_{ref}=10$ and $\Lambda_B = 150/\pi$ Mm. (Bottom) Alfv\'en speed as a function of $y$.}
\end{figure}

\noindent In addition, $B_{ref}$ must satisfy the condition $c_{A}(y_{ref})=10c_{s}(y_{ref})$, and once again we set $B_{ref}=14.472$ G as in the case of magnetic arcades. The resulting magnetic field and Alfv\'en speed are shown in Fig. \ref{fig:Vector_field_funnel_CA}. Due to the symmetries, unlike the arcade, in this case the Alfv\'en speed depends only on $y$.

\subsection{Plasma temperature model}

The temperature field is assumed to obey the C7 model \cite{Avrett&Loeser2008}. This is a semiempirical model of the chromosphere, with the temperature distribution adjusted to obtain optimum agreement between calculated and observed continuum intensities, line intensities, and line profiles of the SUMER \cite{Curdt_et_al_1999} atlas of the extreme ultraviolet spectrum. 

\begin{figure}
\includegraphics[width=6cm]{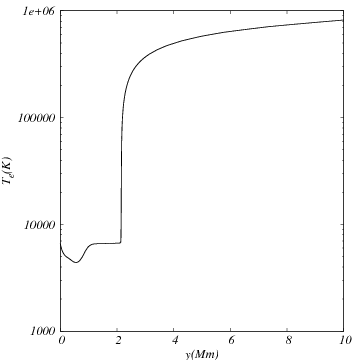}
\caption{\label{fig:Te_C7} Equilibrium temperature profile for the C7 model interpolated into our numerical domain.}
\end{figure}

We set the temperature profile by interpolation of the data in \cite{Avrett&Loeser2008} into our numerical domain, and the result is shown in Fig. \ref{fig:Te_C7}. The main properties of this profile are that $T$ varies nearly three orders of magnitude across the transition between the chromosphere an the corona, ranging from $5000$ K at $y=1.5$ Mm up to $10^6$ K at $y=10$ Mm. For bigger $y$ the temperature profile is assumed to be constant.

\begin{figure}
\includegraphics[width=6cm]{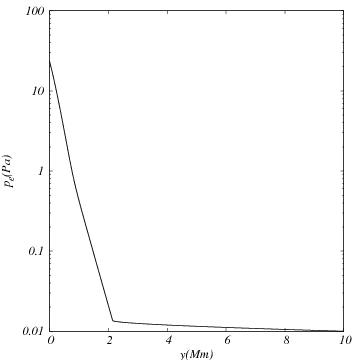}
\includegraphics[width=6cm]{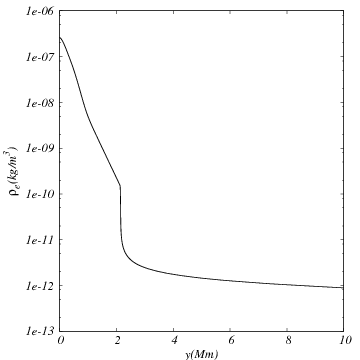}
\caption{\label{fig:rhoe_pe} Equilibrium gas pressure (top) and mass density (bottom) for the C7 model.}
\end{figure}

The pressure $p_{e}(y)$ is integrated numerically using equation (\ref{pe_profile}) and $\rho_{e}(y)$ is calculated using equation 
(\ref{rhoe_profile}). The two profiles are shown in Fig. \ref{fig:rhoe_pe}, where the important gradients at the transition region can be seen.

\subsection{Initial conditions}

Using the previous experience in 
\cite{Murawski&Musielak_2010,Chmielewski_et_al_2013,Woloszkiewicz_et_al_2014,Chmielewski_et_al_2014a,Chmielewski_et_al_2014b} the Alfv\'en waves are considered to be perturbations of the $z-$component of the velocity field $v_{z}$ with a given profile. In our cases, both for the arcades and the funnels we use the velocity profile in \cite{Chmielewski_et_al_2014a}  considering a Gaussian profile of the form

\begin{equation}
v_{z}(x,y,t=0)=A_{v}e^{-\frac{(x-x_{0})^{2}}{w_{x}^{2}}-\frac{(y-y_{0})^{2}}{w_{y}^{2}}}, \label{Gaussian_pulse} 
\end{equation}

\noindent where $A_{v}$ is the amplitude of the pulse, ($x_{0}$, $y_{0}$) is its initial position, $w_{x}$ and $w_{y}$ denote the width of the pulse along $x$ and $y$.  

In this paper we consider the interaction between two pulses, one launched at initial time and a second pulse launched after a given delay lapse $\tau$. We study the effects of the value of the delay lapse on the dispersion of the total energy of the pulses and their capability of transporting energy toward the corona.

\subsection{Numerical methods}

The equations (\ref{dimensionless_density})-(\ref{dimensionless_divergenceB}) are solved numerically using our own numerical code, which is the Newtonian version of CAFE \cite{cafe}, that solves the MHD equations on a uniform cell centered mesh in unigrid mode, and has been applied in relativistic scenarios including the accretion on black holes \cite{grupo2014,grupo2013,grupo2012,grupo2011}. The integration in time uses the method of lines with a third order Runge-Kutta time integrator \cite{Shu&Osher_1989}. The right hand sides of the MHD equations 
are discretized using a finite volume approximation, together with a High Resolution Shock Capturing method. In all the simulations in this paper, this method uses the HLLE \cite{Harten_et_al_1983} approximate Riemann solver formula in combination with the second order linear piecewise reconstructor MINMOD.  

The numerical evolution of initial data involving Maxwell equations leads to the violation of the divergence free constraint  equation (\ref{divergenceB}), developing as a consequence unphysical results like the presence of a magnetic net charge. Among the several methods available to control the growth of the constraint violation \cite{Toth_2000}, in our code we use a version of the constrained transport method (CT) proposed in \cite{Evans&Hawley_1988}, which is based on the use of the fluxes computed with the
conservation scheme itself. This algorithm is known as flux-CT \cite{Balsara_2001}, which maintains the constraint small. 

We use a 2D domain, and the boundary conditions consist in fixing all the evolution variables at all four boundaries to their equilibrium initial values.


\subsection{Alfv\'en waves and coronal heating}

In order to estimate the influence of Alfv\'en waves as carriers of energy to the solar corona, which eventually would produce the heating of the quiet corona or the acceleration of the solar wind, it is necessary to have flux estimates toward the corona. One way to measure the flux energy carried by Alfv\'en waves is to consider the Poynting flux given by 

\begin{equation}
{\bf F}_P = \frac{1}{\mu_0}{\bf B} \times ({\bf v} \times {\bf B})
\end{equation}

\noindent from which we only take the $y-$component $F_{Py}$, that carries energy toward the upper shells.
Another usual measurement involves the WKB approximation, which approximates the total flux formula as \cite{Browning_1991,Hollweg_1991,Kudoh&Shibata_1999}:

\begin{equation}
F \approx \rho\ v_{z}^{2}c_{A}, \label{Energy_Flux_WKB} 
\end{equation}

\noindent where $\rho$ is the density, $c_{A}$ is the local speed of the Alfv\'en waves and $v_{z}$ is the Alfv\'en speed, measured at a particular height.  Here we measure both fluxed in order to compare.

A way to estimate whether the Alfv\'en waves contribute or not to heat the quiet corona is using the threshold in  \cite{Withbroe&Noyes_1977}, indicating that when the flux exceeds the value $\sim$ 3$\times 10^{5}$ erg/s/cm$^2$, the contribution to heating is considerable. Usually what is done is that $F_{WKB}$ is measured at a specific point \cite{Kudoh&Shibata_1999,Chmielewski_et_al_2014b}. Here we measure both fluxes not at a single point, but in order to have a bigger picture we scan these values in three different points specified for each case.


\section{Results of numerical simulations}
\label{sec:results}

We present the results of the energy flux transfered to the corona due to the interaction of two waves evolving on arcades and on funnels. In all cases we estimate the effects of the time elapsed between the two pulses $\tau$. For this we only consider impulsively generated Alfv\'en waves because their higher amplitude corresponds to a better transport of energy.


\subsection{Interaction between two Alfv\'en waves on a solar magnetic arcade}

We set the simulation domain to [1,29]$\times$[1,15] Mm$^2$, covered with 1200$\times$600 cells. We consider large and short arcades, and the length of the arcades depends on the curvature of the magnetic field lines: 
a) For the large arcades we use $x_{0}=2$ Mm, and the Alfv\'en waves propagate higher up along longer and more curved magnetic field lines, b) for short arcades we set $x_{0}=3$ Mm, and the inclination of the magnetic field lines is much larger and the magnetic field lines are less curved.

\subsubsection{Interaction of two pulses in a large arcade}

\begin{figure}
\includegraphics[width=8cm]{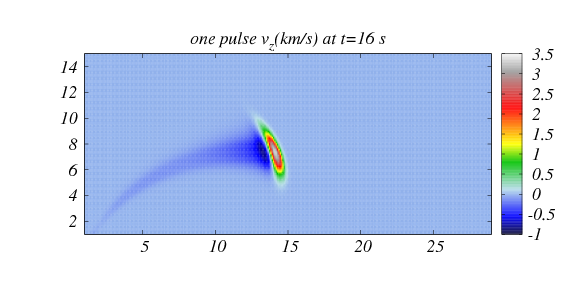}
\includegraphics[width=8cm]{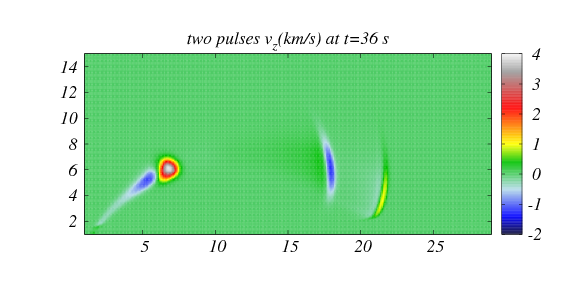}
\includegraphics[width=8cm]{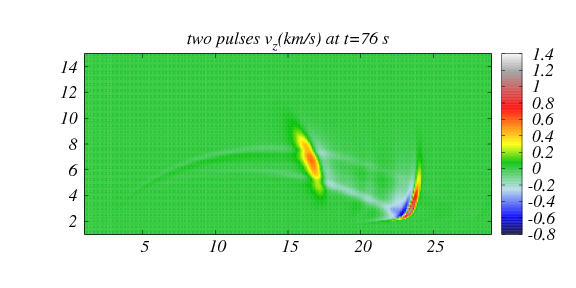}
\includegraphics[width=8cm]{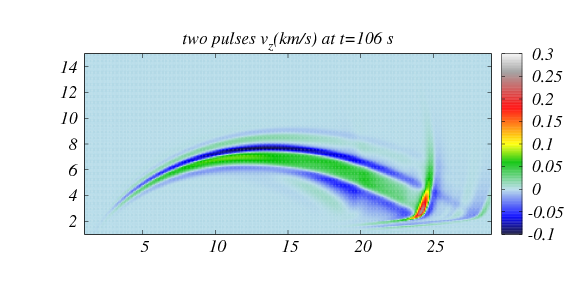}
\includegraphics[width=8cm]{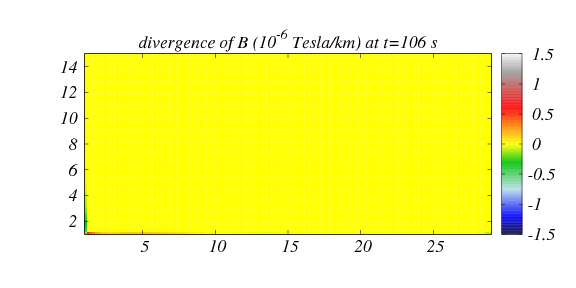}
\caption{\label{fig:Large_arcade_interference_C7} Spatial profiles of the interaction of two Alfv\'en waves on a large arcade with velocity parameters $v_{z}(x,y,t)$, $x_{0}= 2$ Mm, $w_{x}=0.2$ Mm at $t=16$ s, $t=36$ s, $t=76$ s, $t=106$ s. We show the $\nabla\cdot{\bf B}$ at $t=106$ s. This case corresponds to $\tau=30$ s.}
\end{figure}

In this case the amplitude of the initial Gaussian velocity pulse is $A_{v}=3$ km/s, the widths are  $w_{x}=0.2$ Mm, $w_{y}=0.1$ Mm and is located initially at $(x_{0},y_{0})=$(2 Mm, 1.75 Mm) in (\ref{Gaussian_pulse}). We show the results for the particular case in which the second pulse is launched 30 s after the first pulse in Fig. \ref{fig:Large_arcade_interference_C7}. Initially the first pulse decouples into two pulses moving in opposite directions. One pulse propagates downwards very fast, and the second propagates upwards through the chromosphere and transition region to the solar corona, the part of the pulse that propagates upwards reaches the transition region and accelerates due to an increase of the local Alfv\'en speed as expected for the case of a single pulse in \cite{Chmielewski_et_al_2014a}. 
We launch the second pulse at time $t=30$ s, this pulse clearly affects the propagation of the first pulse. The interaction between the two pulses produces reflections as shown at time $t=76$ s. Finally by time $t=106$ s, the waves reach a maximun distance $x$ where they are reflected back to the solar surface. This process shows that phase mixing and reflections grow due to the interaction. Finally, in order to verify our evolution we show at the bottom of Fig. \ref{fig:Large_arcade_interference_C7} the violation of the magnetic field divergence free constraint. 

\begin{figure*}
\includegraphics[width=5cm]{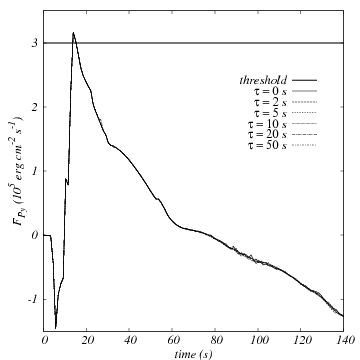}
\includegraphics[width=5cm]{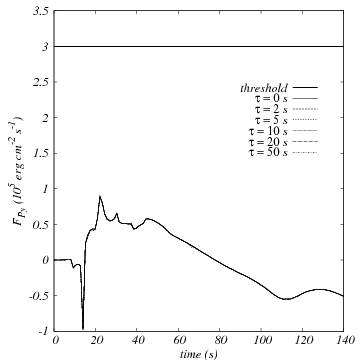}
\includegraphics[width=5cm]{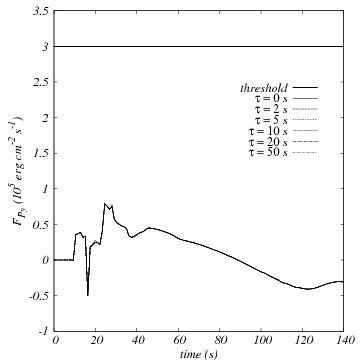}
\includegraphics[width=5cm]{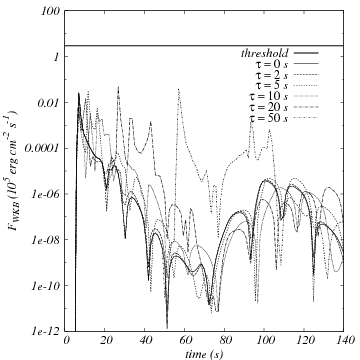}
\includegraphics[width=5cm]{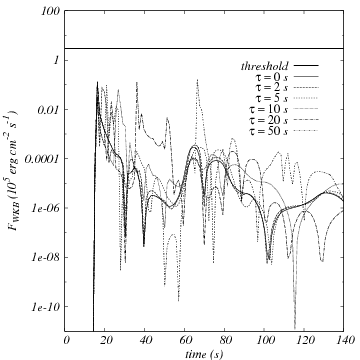}
\includegraphics[width=5cm]{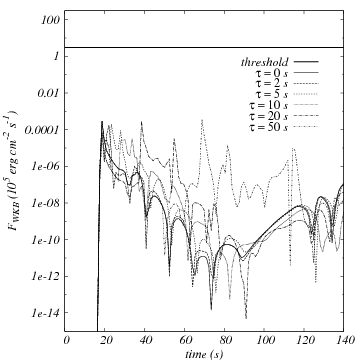}
\caption{\label{fig:Large_arcade_FluxEnergy} Energy fluxes $F_{Py}$ (top) and $F_{WKB}$ (bottom) for the propagation of two pulses in a large arcade using three detectors (from let to right) located at $(7.25,7)$ Mm, $(14.5,7)$ Mm and  $(14.5,10)$ Mm, for $\tau=0,2,5,10,20,50$ s. The continuous line indicates the threshold over which the flux contributes to the coronal heating.}
\end{figure*}

We perform a series of simulations like this one, for various time lapses between the first and the second pulses $\tau$. We measure $F_{Py}$ and $F_{WKB}$ at three different points and show the measurements in Fig.  \ref{fig:Large_arcade_FluxEnergy}. We plot a line indicating the threshold $3 \times 10^{5}$ erg/s/cm$^2$. The results indicate that the Poynting flux measured at the detector located on the left overtakes the threshold during a small lapse of time. The reason is that at this detector the field is more intense. This cannot be observed at other detectors and according to the $F_{WKB}$ the threshold is never achieved.

\subsubsection{Interaction in a short arcade}

\begin{figure}
\includegraphics[width=8cm]{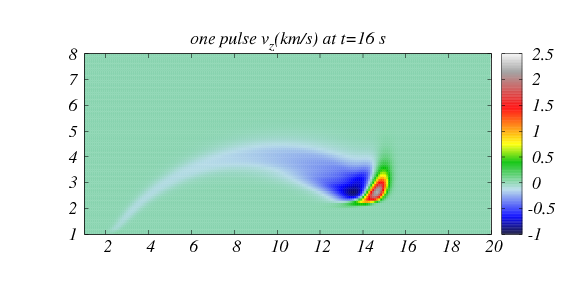}
\includegraphics[width=8cm]{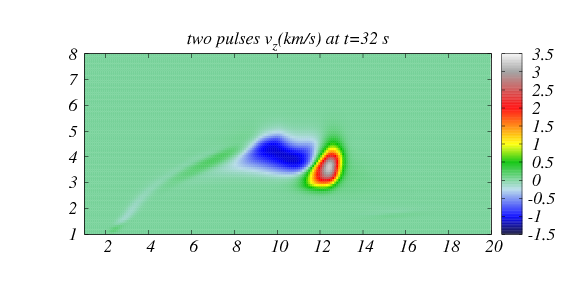}
\includegraphics[width=8cm]{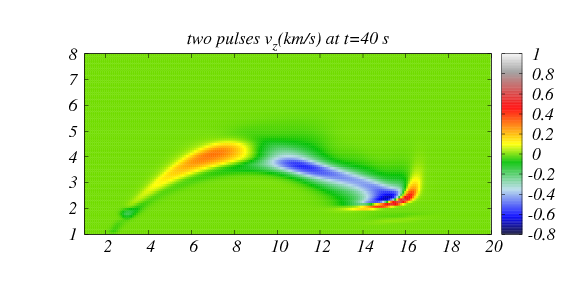}
\includegraphics[width=8cm]{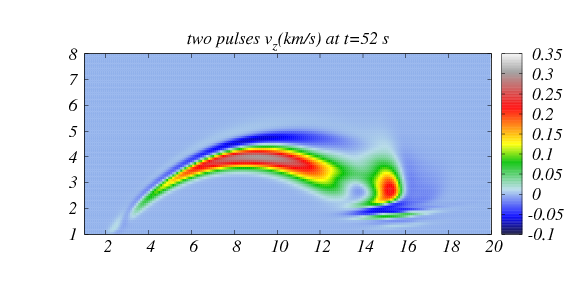}
\caption{\label{fig:Short_arcade_interference_C7} Spatial profiles of the interaction of two Alfv\'en waves $v_{z}(x,y,t)$ for $x_{0}= 3$ Mm, $w_{x}=0.2$ Mm at $t=16$ s, $t=32$ s, $t=40$ s, $t=52$ s.}
\end{figure}

In a similar way, we study the propagation of the same velocity Gaussian pulses with $A_{v}=3$ km/s, $w_{x}=0.2$ Mm, $w_{y}=0.1$ Mm and $(x_{0},y_{0})=$(3 Mm, 1.75 Mm) on a short arcade.
In Fig. \ref{fig:Short_arcade_interference_C7} we show a particular case with $\tau=20$ s. Again, the initial pulse decouples into two pulses moving in opposite directions. At time $t=16$ s we only show the part of the pulse that moves upwards without interaction. This pulse is affected by the transition region and phase mixing, which produces the appearance of a fanning with negative amplitude. We launch the second pulse at time $t=20$ s, which produces interference at time $t=32$ s. By $t=40$ s the second pulse suffers the effect of the density decay in the transition region, and the fanning becomes elongated. At time $t=52$ s the waves reach a maximum distance in $w$, where they suffer strong reflections. Part of these waves return to the solar surface, so as the energy.

\begin{figure*}
\includegraphics[width=5cm]{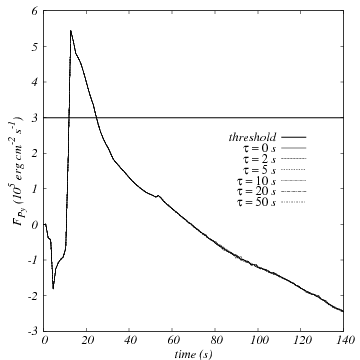}
\includegraphics[width=5cm]{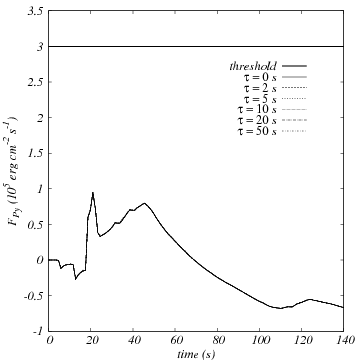}
\includegraphics[width=5cm]{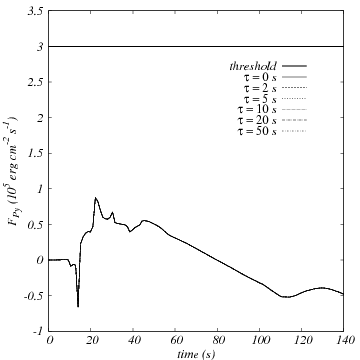}
\includegraphics[width=5cm]{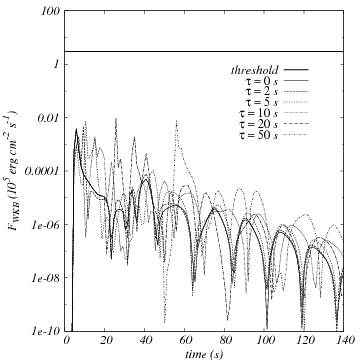}
\includegraphics[width=5cm]{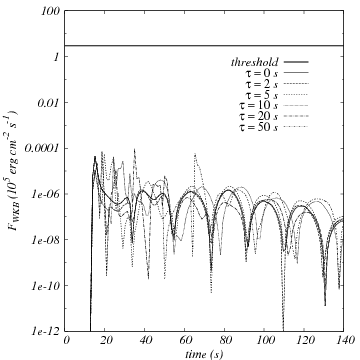}
\includegraphics[width=5cm]{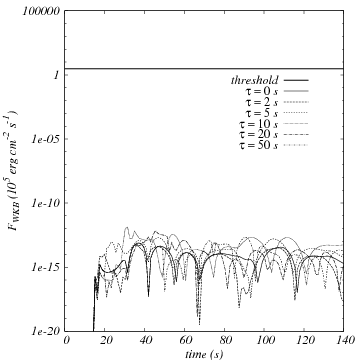}
\caption{\label{fig:Short_arcade_FluxEnergy} Energy fluxes $F_{Py}$ and $F_{WKB}$ for the propagation of two pulses in a large arcade using three detectors at $(7.25,4.5)$ Mm, $(14.5,4.5)$ Mm and  $(14.5,7)$ Mm, for $\tau=0,2,5,10,20,50$ s. Again we show the threshold.}
\end{figure*}

We perform the same experiments as in the previous case for various time lapses between the first and the second pulses. Again we use three detectors to measure $F_{Py}$ and $F_{WKB}$ shown in Fig. \ref{fig:Short_arcade_FluxEnergy}. Like in the previous case, only the Poynting flux reports values above the threshold in the detector at the left.

\subsection{Interaction between two Alfv\'en waves in a coronal funnel}

In this case we set the simulation domain to be [-5,5]$\times$[1, 81] Mm$^2$, covered with 300$\times$2400 cells. Like in the cases of arcades, we launch two velocity pulses with profile (\ref{Gaussian_pulse}) separated by an interval of time $\tau$ on the funnel magnetic field configuration defined in \ref{sec:funnels}.

\subsubsection{Interaction between two Alfv\'en waves with amplitude $A_{v}=10$ km/s}

\begin{figure*}
\includegraphics[width=4.cm]{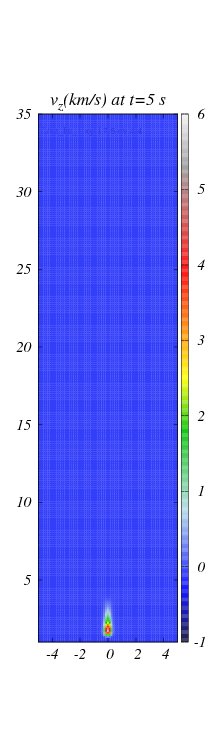}
\includegraphics[width=4.cm]{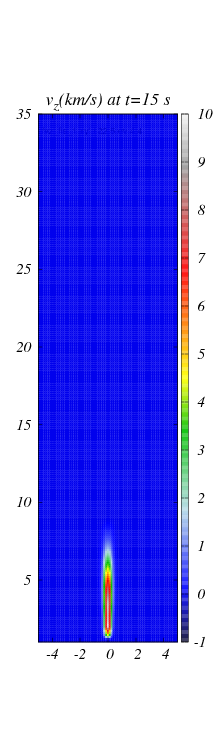} 
\includegraphics[width=4.cm]{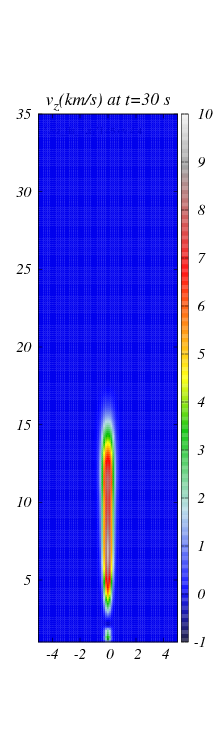}
\includegraphics[width=4.cm]{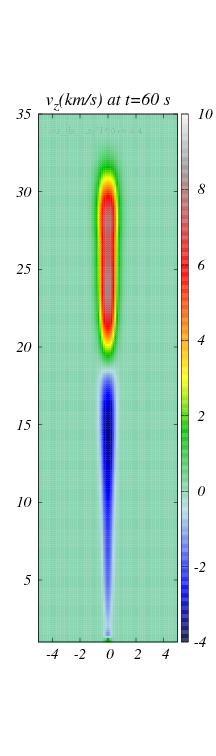}
\caption{\label{fig:vz_Av10_interference_tpulse10} Spatial profiles of the interaction of two non-linear pulses $v_{z}(x,y,t)$ for $A_{v}=10$ km/s, $w_{x}=w_{y}=0.25$ Mm displayed at times $t=5$ s, $t=15$ s, $t=30$ s, $t=60$ s. In this case $\tau=10$ s.}
\end{figure*}

The typical behavior of the pulses is shown in Fig. \ref{fig:vz_Av10_interference_tpulse10} for $\tau=10$ s and pulses with parameters $A_{v}=10$ km/s, $w_{x}=w_{y}=0.2$ Mm, ($x_{0}=$0, $y_{0}=$1.75) Mm. Initially the first pulse decomposes into two propagating waves \cite{Chmielewski_et_al_2014b}, one pulse that moves upwards and a second pulse that moves downwards. The amplitude of the pulse that moves downwards decays with time and practically disappears. We show the first pulse at time $t=5$ s, this pulse experiences an acceleration at the transition region due to an increment of the local Alfv\'en speed $c_{A}$. The shape of the pulse becomes ellipsodial elongated along the vertical direction. Then the second Gaussian pulse appears at time $t=10$ s. Still by $t=15$ s the effect of the second pulse is not significant to the shape of the first pulse. However, at that time the pulse that moves upwards suffers another elongation. At later times in the evolution, we can see reflections and phase mixing because of the effect of the transition region in the pulses that move upwards at time $t=30$ s. The two pulses have already reached the solar corona. At time $t=60$ s, a signal of $v_{z}$ with negative amplitude reaches the solar corona, and the pulses that move upwards continue moving vertically.

\begin{figure*}
\includegraphics[width=5cm]{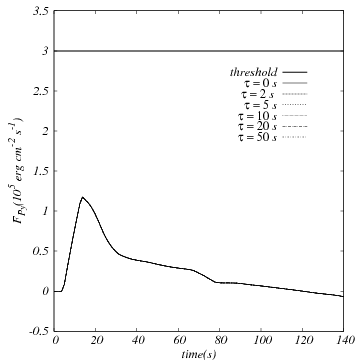}
\includegraphics[width=5cm]{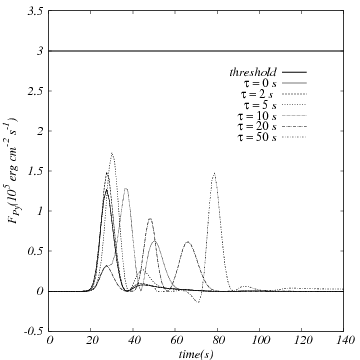}
\includegraphics[width=5cm]{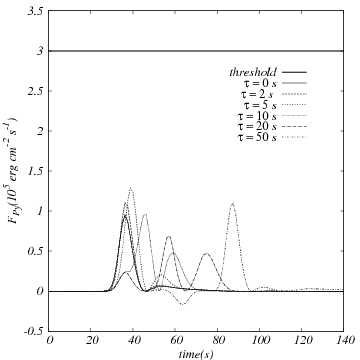}
\includegraphics[width=5cm]{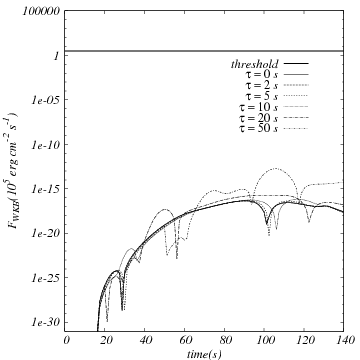}
\includegraphics[width=5cm]{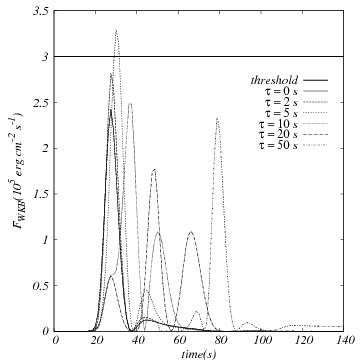}
\includegraphics[width=5cm]{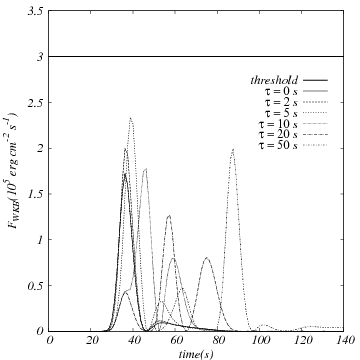}
\caption{\label{fig:Funnel_A10_FluxEnergy} Energy fluxes $F_{Py}$ and $F_{WKB}$ for the propagation of two pulses in a coronal funnel using three detectors at $(-2.5,10)$ Mm, $(0,10)$ Mm and  $(0,15)$ Mm, for $\tau=0,2,5,10,20,50$ s. The threshold is achieved only for a particular combination $\tau=5$s, however this is not a generic case.}
\end{figure*}

The fluxes measured at various detectors are shown in Fig. \ref{fig:Funnel_A10_FluxEnergy}.
In this case the Poynting flux never reaches the threshold for any of the combination of pulses whereas, unlike the arcade cases, the $F_{WKB}$ flux exceeds the threshold during a small lapse of time for the case $\tau=5$s, measured by a detector located along the central vertical line (0,10).

\subsubsection{Interaction between two Alfv\'en waves with amplitude $A_{v}=40$ km/s}

\begin{figure*}
\includegraphics[width=4.cm]{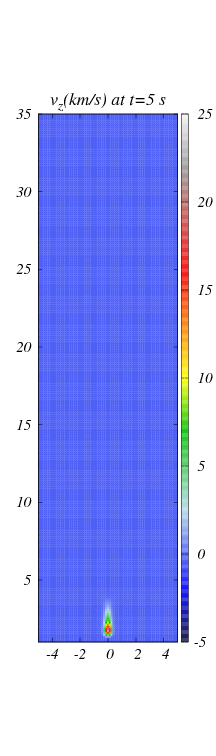}
\includegraphics[width=4.cm]{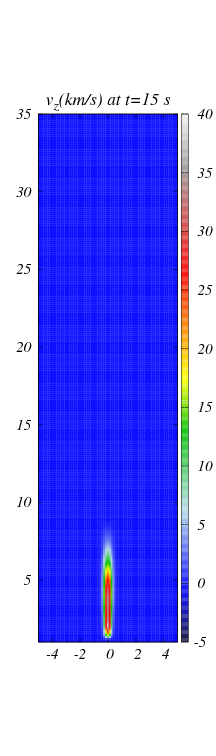} 
\includegraphics[width=4.cm]{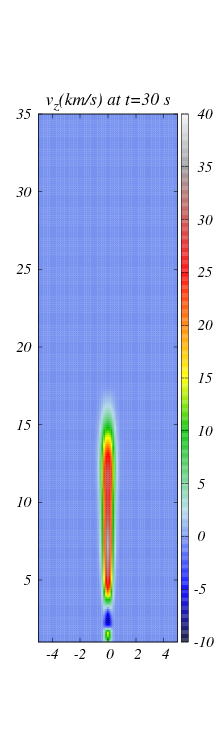}
\includegraphics[width=4.cm]{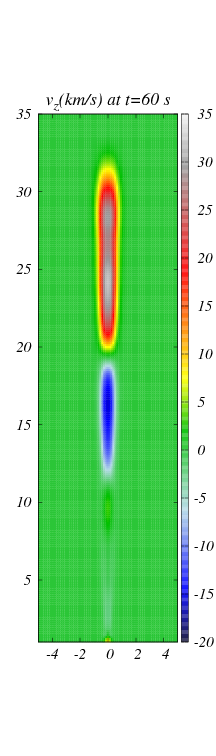}
\caption{\label{fig:vz_Av40_interference_tpulse10} Spatial profile of the interaction of two non-linear pulses $v_{z}(x,y,t)$ for $A_{v}=40$ km/s, $w_{x}=w_{y}=0.2$ Mm displayed at times $t=5$ s, $t=15$ s, $t=30$ s, $t=60$ s. In this case $\tau=10$ s.}
\end{figure*}

In order to show the contribution using one of  the highest amplitudes found in literature, we prepare a case of an impulsively generated and high amplitude pulse  with parameters $A_{v}=40$ km/s, $w_{x}=w_{y}=0.2$ Mm, ($x_{0}=$0, $y_{0}=$1.75) Mm \cite{Chmielewski_et_al_2013,Chmielewski_et_al_2014b}.
We show snapshots of the evolution of the wave in Fig. \ref{fig:vz_Av40_interference_tpulse10}.

\begin{figure*}
\includegraphics[width=5cm]{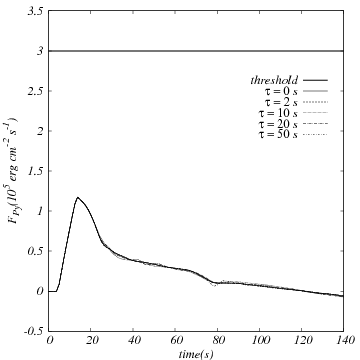}
\includegraphics[width=5cm]{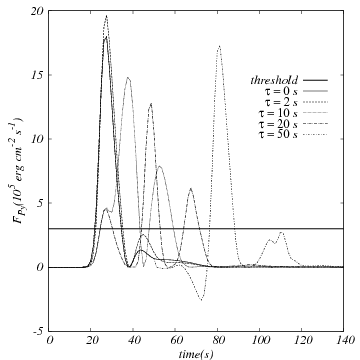}
\includegraphics[width=5cm]{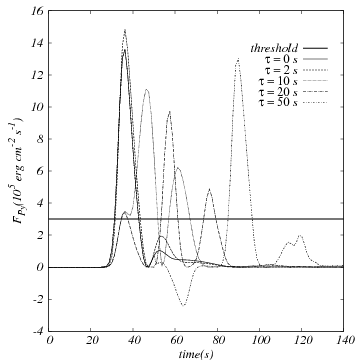}
\includegraphics[width=5cm]{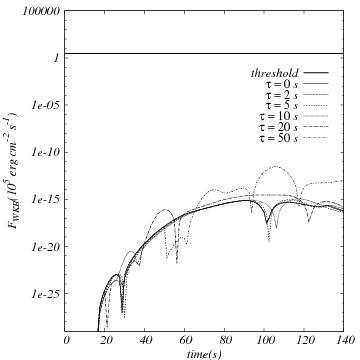}
\includegraphics[width=5cm]{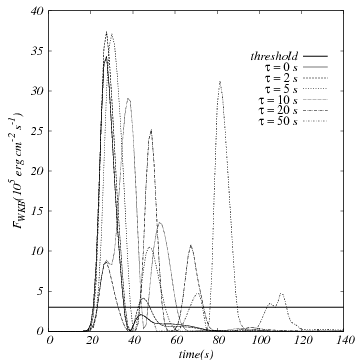}
\includegraphics[width=5cm]{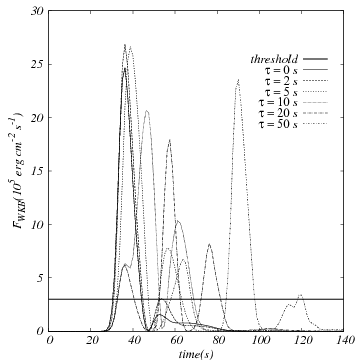}
\caption{\label{fig:Funnel_A40_FluxEnergy} Energy fluxes $F_{Py}$ and $F_{WKB}$ for the propagation of two pulses in a coronal funnel using three detectors at $(-2.5,10)$ Mm, $(0,10)$ Mm and  $(0,15)$ Mm, for $\tau=0,2,5,10,20,50$ s. The threshold is overtaken in a systematic way for all the values of $\tau$ used.}
\end{figure*}

In Fig. \ref{fig:Funnel_A40_FluxEnergy} we show $F_{Py}$ and $F_{WKB}$ at various detectors. We find that the two detectors located at the center of the domain measure fluxes, both $F_{Py}$ and $F_{WKB}$ over the threshold for various values of $\tau$ in a generic way.

As mentioned before, this is one of the highest value used to model the propagation of Alfv\'en waves, however it is not clear from observations that such high speeds have been observed and associated to funnel type of configurations so far. Therefore we have only shown that in the case that such high speeds are observed, they may actually contribute to heating the corona.

\section{Conclusions}
\label{sec:conclusions}

We have proposed an optimistic scenario where not only a single Alfv\'en wave pulse can contribute to heating the quiet corona, but two possible pulses that may be launched very close in time.

We found that impulsively generated linear Alfv\'en waves in the solar magnetic arcades, with slow pulses, in large or short arcades, carry enough energy as to heat the solar quiet corona only during a small lapse of time for a very particular combination of pulses, which indicates the threshold is not overtaken for generic combination of pulses.

In the case of the propagation of impulsively generated non-linear Alfv\'en waves in the coronal funnels, the waves with amplitud $A_{v}=10$ km/s do not carry enough energy flux as to heat the solar quiet corona, even if two consecutive pulses are launched. On the other hand, the energy transferred overtakes the threshold by orders of magnitude in the case of waves with an amplitude of $A_{v}=40$ km/s, which we cannot be sure that exist but has been mentioned in existing literature.

We conclude from our analysis, that impulsively generated Alfv\'en waves, even in this optimistic scenario, are unlikely candidates to heat the quiet corona.


\section*{Acknowledgments}

We thank Victor De la Luz and Ernesto Aguilar for reading the manuscript and providing important comments. This research is partly supported by grant CIC-UMSNH-4.9.


\bsp

\label{lastpage}

\end{document}